\documentclass[11pt,a4paper]{article}
\usepackage[utf8]{inputenc}
\usepackage[T1]{fontenc}
\usepackage{lmodern}
\usepackage{hyperref}
\hypersetup{hidelinks}
\usepackage{graphicx}
\usepackage{listings}
\usepackage{xcolor}
\usepackage{booktabs}
\usepackage{amsmath}
\usepackage{geometry}
\usepackage{natbib}
\usepackage{fancyhdr}
\usepackage{float}
\usepackage{subcaption}
\usepackage{multirow}

\definecolor{codegreen}{rgb}{0,0.6,0}
\definecolor{codegray}{rgb}{0.5,0.5,0.5}
\definecolor{codepurple}{rgb}{0.58,0,0.82}
\definecolor{backcolour}{rgb}{0.95,0.95,0.92}

\lstdefinestyle{pythonstyle}{
    backgroundcolor=\color{backcolour},   
    commentstyle=\color{codegreen},
    keywordstyle=\color{magenta},
    numberstyle=\tiny\color{codegray},
    stringstyle=\color{codepurple},
    basicstyle=\ttfamily\footnotesize,
    breakatwhitespace=false,         
    breaklines=true,                 
    captionpos=b,                    
    keepspaces=true,                 
    numbers=left,                    
    numbersep=5pt,                  
    showspaces=false,                
    showstringspaces=false,
    showtabs=false,                  
    tabsize=2,
    language=Python
}

\title{\textbf{AMBER: A Columnar Architecture for Scalable Agent-Based Modelling in Python}}

\author{Anh-Duy Pham\\
\small University of Würzburg, Germany\\
\small \texttt{ORCID: 0000-0003-3832-9453}
}

\date{May 2026}

\begin{document}

\maketitle

\begin{abstract}
Python is widely used for agent-based modelling because it is accessible and has a mature scientific ecosystem, but object-per-agent execution incurs interpreter overhead that restricts the population sizes feasible in interactive modelling, calibration, and parameter sweeps. This paper presents AMBER, a Python framework that stores agent state in a Polars-backed columnar table and exposes population operations through a compact view API. The framework preserves conventional model and agent abstractions while translating common population updates into compiled column operations; behaviours that do not vectorise remain expressible through a buffered object-oriented path. We evaluate AMBER on wealth transfer, random walk, and spatial SIR benchmarks against Mesa, AgentPy, SimPy, Melodie, Agents.jl, and AMBER's own loop path, using invariant checks to verify comparable model outputs before timing. Across the tested workloads, AMBER has the lowest execution time among Python-hosted implementations and achieves speedups of up to 1118$\times$ over Mesa; on the largest SIR benchmark it is also faster than the Julia-based Agents.jl implementation.
\end{abstract}

\vspace{1em}

\section{Introduction}

Agent-based modelling (ABM) studies complex adaptive systems by simulating autonomous agents whose local interactions generate emergent macro-level phenomena \citep{abar2017agent}. The methodology is central to computational epidemiology, ecology, economics, and social simulation, where it captures agent heterogeneity, spatial structure, and adaptive behaviour in ways that aggregate equation-based models cannot.

Python is widely used for ABM in the research community, primarily through Mesa \citep{masad2015mesa} and AgentPy \citep{foramitti2021agentpy}. Their accessibility has enabled broad adoption, but both, along with most Python-hosted ABM tools, share an architectural assumption that limits scalability: each agent is a Python object, and population-wide updates are Python-level loops over those objects. For populations of thousands to tens of thousands of agents, the interpreter overhead of this pattern can exceed the cost of the model-specific computation, constraining parameter sweeps, Monte Carlo replicates, and model complexity.

AMBER addresses this bottleneck through two architectural choices. First, agent state is centralised in a columnar DataFrame backed by Polars \citep{polars_zenodo}, a Rust-based data processing library. Second, a \emph{view API} supports full-population, filtered, and scatter updates by compiling user-facing expressions down to a small number of Polars operations. Model code remains in Python while computational kernels execute in compiled code. A conventional object-oriented \texttt{Agent} class remains available for per-agent logic that does not vectorise.

This paper investigates whether columnar state management can serve as a practical foundation for large-scale ABM in Python. The central question is not only whether columnar storage is faster for favourable kernels, but whether it can be integrated into a complete modelling framework without abandoning the abstractions that make Python ABM accessible.

\subsection{Research Questions and Contributions}

The work is organised around three research questions:

\begin{itemize}
\item \textbf{RQ1: Architectural viability.} Can columnar state management support ABM semantics while preserving familiar model, agent, and population abstractions?
\item \textbf{RQ2: Performance characterisation.} Which workload characteristics benefit from columnar execution, and where does a Python-hosted columnar framework remain limited by call overhead or non-vectorisable logic?
\item \textbf{RQ3: Framework completeness.} Can a columnar architecture support the surrounding infrastructure required for practical ABM work, including spatial environments, experiment management, optimisation, and compatibility with per-agent logic?
\end{itemize}

This paper makes four contributions. First, it formalises columnar state management as an architectural alternative to object-per-agent representation in Python-hosted ABM. Second, it presents AMBER, an implementation of this architecture with a view API for full-population, filtered, and id-indexed scatter updates. Third, it provides a correctness-checked benchmark suite comparing AMBER with six alternative implementations, correcting cross-framework mismatches before reporting performance. Fourth, it characterises the regimes in which columnar execution provides substantial advantages and the regimes in which compiled per-agent loops remain preferable.

\subsection{Scope and Positioning}

Researchers who build large ABMs often face a trade-off between accessibility and execution speed. Remaining in Python preserves the surrounding scientific workflow but inherits the interpreter-overhead ceiling of Mesa or AgentPy. Moving to Java, Julia, or GPU-oriented C++ can improve throughput, but it introduces additional language, tooling, and maintenance requirements around models that are otherwise primarily scientific artifacts.

AMBER is designed for this intermediate design point: Python source code, familiar ABM abstractions, and compiled execution for population-wide operations. Models expressed in Mesa or AgentPy can be translated with limited structural change because AMBER preserves the conventional model, agent, and agent-list hierarchy. The principal modelling change is that population-wide updates are expressed as views over the population rather than loops over individual agents.

\section{Related Work}

\subsection{Python ABM Frameworks}

Mesa \citep{masad2015mesa} established the prevalent Python ABM pattern: agents as Python class instances organised by a scheduler. It introduced widely-adopted idioms including datacollectors for state recording and modular space classes. Its design prioritises accessibility and flexibility, making it a common teaching and prototyping framework.

AgentPy \citep{foramitti2021agentpy} refines Mesa's approach with improved ergonomics, tighter Jupyter integration, and a built-in parameter-sweep API. Its \texttt{AgentList} exposes aggregate operations over collections, though the underlying storage is still object-per-agent.

Melodie \citep{yu2023melodie} separates models into agents, environment, scenario, and calibrator components, accelerating selected execution paths through Cython compilation. ABSESpy \citep{wang2024absspy} extends Mesa with domain-specific abstractions for social-ecological systems modelling. SimPy \citep{zinoviev2024discrete} is a discrete-event simulation library rather than a dedicated ABM framework, but it is included here as a Python baseline because it is sometimes repurposed for agent simulation.

\subsection{Compiled and Accelerated ABM Frameworks}

Repast \citep{north2013complex} and MASON \citep{luke2005mason} are mature Java-based platforms offering sophisticated scheduling, GIS integration, and checkpointing. Agents.jl \citep{datseris2024agents} leverages Julia's JIT compilation and multiple dispatch to combine scripted development with compiled execution; it is the direct non-Python performance baseline in our evaluation. FLAME GPU 2 \citep{richmond2023flame} maps agent operations onto CUDA kernels for GPU-accelerated simulation at massive scale but requires CUDA expertise and GPU-amenable model structure. NetLogo \citep{tisue2004netlogo} remains influential for its visual development environment and domain-specific language.

\subsection{Columnar Data Processing}

Columnar storage---organising data by columns rather than rows---has revolutionised analytical database systems and scientific data processing \citep{arrow2026}. Modern CPUs favour contiguous memory access, SIMD instructions process multiple values per cycle, and column operations parallelise naturally. Polars \citep{polars_zenodo}, built on Apache Arrow's columnar memory format and implemented in Rust, exposes these capabilities through a Python-accessible DataFrame API with multi-threaded execution. AMBER's contribution is applying these database-oriented techniques to the distinct access patterns of agent-based simulation and hiding them behind an ABM-native abstraction.

\section{Architectural Design}
\label{sec:architecture}

AMBER is organised around a columnar state store, a user-facing view API, and conventional ABM components for environments, experiments, and optimisation. The design goal is to expose vectorised execution as a first-class modelling idiom while retaining the object-oriented abstractions used by existing Python ABM frameworks.

\subsection{The Paradigm Shift: From Objects to Columns}

Conventional ABM frameworks represent each agent as a Python object on the heap, with attributes stored in the object's \texttt{\_\_dict\_\_} and the population held as a list or dictionary of such objects. Memory access is therefore indirect: reading a single attribute across $N$ agents requires $N$ object lookups and $N$ dictionary accesses. A population-wide update such as incrementing every agent's wealth becomes $N$ bytecode-interpreted iterations, each paying Python's overhead for reference counting, type checking, and attribute access. For simple attribute updates, this interpreter overhead can dominate the model-specific computation.

AMBER replaces this representation with a single Polars DataFrame in which each agent attribute occupies a contiguous typed column, and agents are identified by a sequential integer id. A population-wide read traverses one typed array; a population-wide write lowers to a vectorised Polars operation. Agent state is therefore maintained in the DataFrame as the single source of truth, while Python-level \texttt{Agent} instances, when used, act as row-oriented accessors rather than independent state containers.

\begin{figure}[H]
    \centering
    \includegraphics[width=0.9\textwidth]{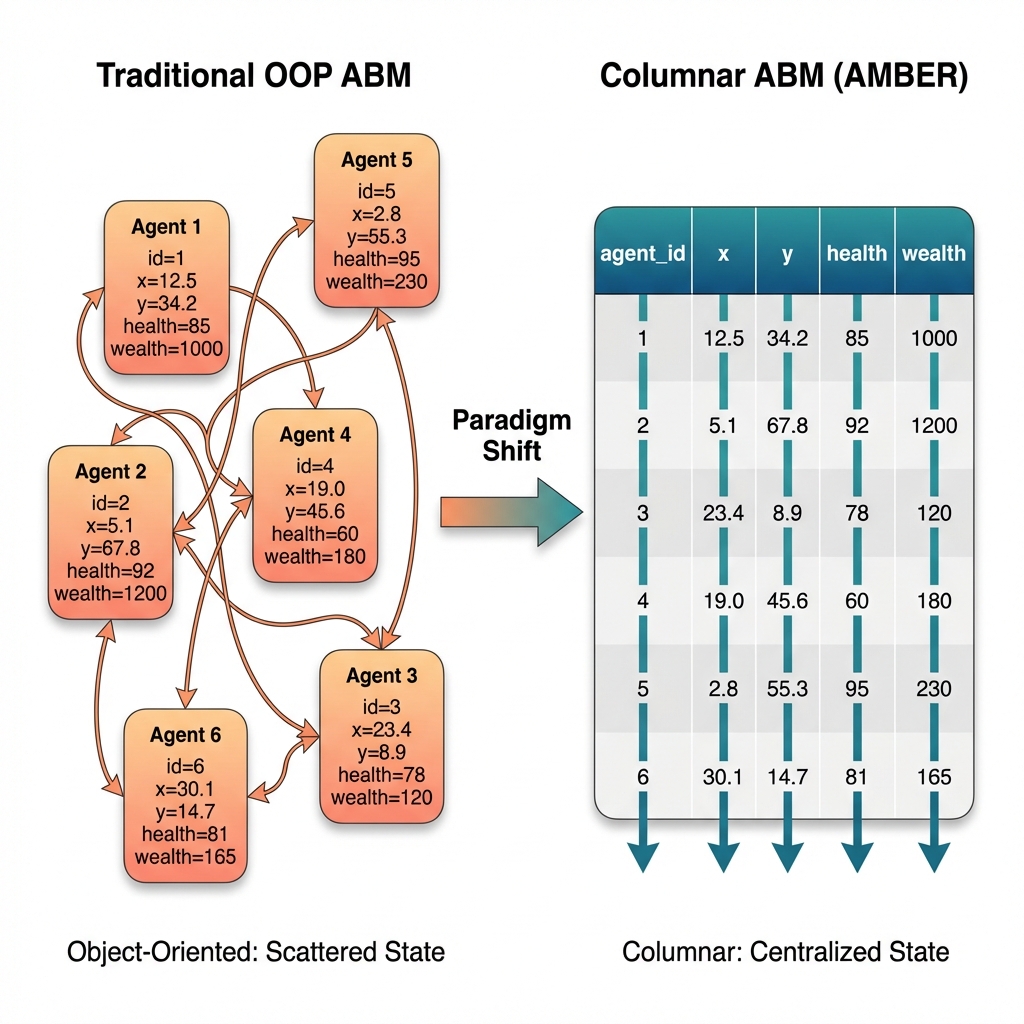}
    \caption{State-management contrast between object-oriented ABM and AMBER's columnar representation. Object-per-agent frameworks distribute attributes across independent Python objects, whereas AMBER stores attributes as typed columns in a shared table. The columnar layout enables population-wide operations to execute as compiled array transformations.}
    \label{fig:paradigm}
\end{figure}

\subsubsection{Columnar State Management}

All agent state resides in a single \texttt{Population} object that wraps a Polars DataFrame where each attribute is a column. Column data is stored in contiguous typed arrays, with rows corresponding to agents identified by sequential integer IDs.

A population-wide update becomes a single vectorized operation expressed through AMBER's \emph{view API}:

\begin{lstlisting}[caption={Vectorized population-wide update via the view API}]
# Increment wealth for every agent in a single operation
self.agents.wealth += 10
\end{lstlisting}

This assignment invokes a single Python-to-Rust call. The attribute read returns a Polars Series sourced from the population DataFrame; the arithmetic executes in compiled Rust, and the assignment lowers to one \texttt{df.with\_columns(...)} call that replaces the column in place. For subset updates, the same syntax extends naturally:

\begin{lstlisting}[caption={Subset-conditional update via a filtered view}]
# Only agents with positive wealth give away one unit
donors = self.agents.where(self.agents.wealth > 0)
donors.wealth -= 1
\end{lstlisting}

The \texttt{where} call returns a \texttt{FilteredAgentList} view keyed by the ids matching the predicate; the subsequent assignment applies the update to exactly those rows via a single hash join against the population DataFrame. The update therefore avoids a Python-level loop over individual agents.

For id-indexed updates where an agent may appear multiple times (for example, scattering resources from $k$ donors to $k$ randomly chosen recipients), AMBER provides a \texttt{scatter\_add} primitive:

\begin{lstlisting}[caption={Id-indexed accumulate with duplicate handling}]
# Each donor picks a random recipient; credits accumulate correctly
# when two donors happen to choose the same agent.
recipients = self.nprandom.choice(self.agents.ids.to_numpy(),
                                  size=len(donors))
self.agents.at[recipients].scatter_add(wealth=1)
\end{lstlisting}

\texttt{scatter\_add} uses a numpy \texttt{np.add.at} implementation with a cached id-to-row-position lookup, handling duplicate ids correctly by summing their contributions rather than overwriting them, while executing in a single pass over the affected rows.

\subsection{Performance Model}

Let $T_{OOP} = N \cdot (c_{iter} + c_{dict} + c_{op})$ be the cost of a population-wide attribute update in an object-oriented framework, where $c_{iter}$ and $c_{dict}$ are Python iteration and dictionary-access overheads and $c_{op}$ is the operation-specific computation. Let $T_{col} = c_{call} + (N/k) \cdot c_{op}$ be the corresponding cost in a columnar framework, where $c_{call}$ is the fixed cost of invoking a vectorised operation and $k$ is the effective SIMD parallelism factor. For large $N$ and low-arithmetic-intensity updates, the expected speedup is $S \approx k \cdot (c_{iter} + c_{dict}) / c_{op}$. The model therefore predicts that columnar execution is most beneficial when simple operations are applied to large populations. The evaluation in Section~\ref{sec:evaluation} is consistent with this prediction: AMBER's advantage over Mesa ranges from 23$\times$ on random walk to 1118$\times$ on population-wide wealth transfer.

\subsubsection{Agent-Specific Operations}

Not all ABM operations benefit equally from columnar storage. Many SIR-style infection rules, Schelling-style satisfaction checks, and graph-traversal behaviours are naturally expressed as conditional logic over each agent's local neighbourhood. AMBER preserves an explicit object-oriented path for these cases via the traditional \texttt{Agent} class:

\begin{lstlisting}[caption={Per-agent logic using the object-oriented path}]
class InfectedAgent(am.Agent):
    def step(self):
        neighbors = self.get_neighbors(pl.col("status") == "S")
        if neighbors.height > 0 and self.model.random.random() < 0.1:
            # Per-agent record; buffered and flushed in one join.
            neighbors[0].record("status", "I")
\end{lstlisting}

Per-agent \texttt{record} and \texttt{update\_data} calls still use batched storage internally. AMBER's \texttt{Model} class buffers them into a \texttt{\_pending\_writes} dictionary and flushes the entire step's worth of updates via a single \texttt{df.update(on='id')} hash join on the next read of \texttt{agents\_df}. Thus the classical object-oriented \texttt{for agent in model.agents: agent.step()} pattern pays one DataFrame update per step rather than one update per agent.

\subsection{The View API}

A direct columnar implementation would expose DataFrame manipulation throughout model code. AMBER instead provides a \emph{view API} that wraps the underlying Polars DataFrame in three view types, each honouring the same attribute protocol:

\begin{itemize}
\item \textbf{Full-population views} expose every agent in the model.
\item \textbf{Filtered views} expose a predicate-defined subset, such as agents satisfying a Polars expression.
\item \textbf{Scatter views} expose an id-indexed subset where ids may repeat, enabling duplicate-safe accumulation.
\end{itemize}

All three views share the same contract: a column attribute read returns a \texttt{pl.Series} sourced from \texttt{model.agents\_df}, and a column attribute assignment queues through the same batched flush path. The DataFrame is the single source of truth; there is no Python-attribute shadow copy, which prevents inconsistencies between object state and columnar state.

Predicates accept either a boolean Polars Series or a raw Polars expression, so the full Polars expression language remains available without requiring every model to be written directly against the DataFrame API. For example, \texttt{self.agents.wealth > 0} (a Series-returning comparison) and \texttt{pl.col("wealth") > 0} (an expression tree) are both valid arguments to \texttt{where}.

The \texttt{Model} class complements the view API with a columnar bulk-create helper, \texttt{add\_agents}, which initializes $n$ agents in a single call accepting per-column scalar broadcasts, lists, numpy arrays, or Polars Series. This removes repeated single-agent construction from large-population setup.

The \texttt{Population} class remains accessible as the lower-level columnar foundation, exposing \texttt{batch\_update\_by\_ids}, \texttt{batch\_add\_agents}, and \texttt{batch\_update} with a selector expression. Framework extension authors and performance-sensitive users can drop down to this layer; the view API is built on top of it.

\subsection{State Evolution Semantics}

ABM requires mutable state evolution over simulation time. AMBER supports both synchronous and asynchronous update semantics.

In synchronous (simultaneous) updates, all agents act on a consistent state snapshot. This mode enables full vectorization, as entire columns can be transformed simultaneously based on current values without interference between agents.

In asynchronous (sequential) updates, agents observe effects of prior agents' actions within the same time step. This mode is necessary for some model semantics but limits vectorization, as each agent's computation may depend on modifications made by previously processed agents.

\section{Framework Components}
\label{sec:components}

Beyond the core columnar architecture and view API, AMBER provides the infrastructure a full ABM workflow requires.

\subsection{Spatial Environments}

Grid, continuous-space, and network environments share a unified interface. Grid environments implement discrete $n$-dimensional lattices with Moore or von Neumann neighbourhoods and optional torus wrapping; agent positions are stored as integer columns in the Population DataFrame, enabling vectorised neighbour queries. Continuous-space environments provide radius-based neighbour queries and can use a \texttt{scipy.spatial.cKDTree} index for $O(\log N)$ lookups. Network environments integrate NetworkX for graph-based connectivity and ship built-in generators for Erd\H{o}s-R\'{e}nyi, Watts-Strogatz, and Barab\'{a}si-Albert topologies.

\subsection{Experiment Management}

Systematic parameter exploration is central to ABM research because individual simulation trajectories are often stochastic and parameter-sensitive. The \texttt{Experiment} class defines parameter spaces using \texttt{Sample} for discrete values and \texttt{IntRange} for integer ranges, generates Cartesian products over combinations, supports configurable replication counts, and aggregates results directly into columnar form for downstream analysis. \texttt{ParallelRunner} provides multiprocessing-based parallel execution for independent simulations.

\subsection{Optimisation and Calibration}

Model calibration requires efficient search over parameter spaces whose evaluations may be expensive. AMBER includes grid search for low-dimensional discrete spaces, random search for high-dimensional exploration, Bayesian optimisation for sample-efficient calibration, and SMAC integration for multi-fidelity and multi-objective optimisation.

\subsection{Performance Utilities}

AMBER also exposes lower-level utilities for performance-critical components. The spatial index wrapper provides radius and nearest-neighbour queries for continuous-space models. Pre-built vectorised utilities implement recurring patterns such as movement updates, wealth transfers, and random displacement generation, allowing users to reuse tested kernels rather than reimplementing them inside model code.

\section{Empirical Evaluation}
\label{sec:evaluation}

We evaluate AMBER against six other agent-based modelling frameworks. The evaluation combines wall-clock timing with a correctness audit that verifies comparable model outputs before performance results are interpreted.

\subsection{Experimental Design}

\subsubsection{Frameworks Compared}

We compare seven framework implementations. \textbf{AMBER (vectorized)} uses the view API described in Section~\ref{sec:architecture}. \textbf{AMBER (loop)} uses AMBER's object-oriented update path and provides a baseline at the same abstraction level as object-oriented Python ABM frameworks. \textbf{Mesa} \citep{masad2015mesa} and \textbf{AgentPy} \citep{foramitti2021agentpy} are the two mainstream Python ABM frameworks. \textbf{SimPy} \citep{zinoviev2024discrete} is included as a discrete-event alternative with a tight C-based event-loop core. \textbf{Melodie} \citep{yu2023melodie} represents Cython-accelerated Python ABM. \textbf{Agents.jl} \citep{datseris2024agents} provides a Julia JIT-compiled baseline outside the Python ecosystem.

\subsubsection{Benchmark Models}

We selected three models representing distinct computational patterns. The \textbf{Wealth Transfer} model has each agent with positive wealth transfer one unit to a randomly selected other agent at each step. This workload is dominated by attribute reads and writes and represents a low-arithmetic-intensity case for vectorization.

The \textbf{SIR Epidemic} model places agents in continuous 2D space (100$\times$100) where they transition between Susceptible, Infected, and Recovered states based on contact with infected neighbors within a fixed radius (5 units), transmission probability 0.1, and recovery time 14 steps. Five agents begin in the Infected state. This workload involves all-pairs spatial distance computation (quadratic in population size) and state-transition logic.

The \textbf{Random Walk} model has agents move in continuous 2D space using independent uniformly sampled $x$ and $y$ displacements with boundary clamping. This tests position update kernels that are fully vectorizable.

\subsubsection{Correctness Verification}

Performance comparisons are valid only when the compared implementations compute equivalent models. Before any timing runs, we execute each framework against a dedicated correctness script, \path{benchmarks/correctness_check.py}, that checks per-model output invariants:

\begin{itemize}
\item \textbf{Wealth transfer}: total wealth must equal $N \cdot w_{init}$ at any step (Boltzmann conservation).
\item \textbf{Random walk}: all agent coordinates must lie within the configured world bounds (boundary clamp integrity).
\item \textbf{SIR epidemic}: the cardinalities $|S| + |I| + |R|$ must equal $N$ at every step, and the initial infected count must match the configured \texttt{initial\_infected} parameter.
\end{itemize}

This audit identified multiple benchmark defects: SimPy's wealth-transfer helper used a race-prone event update, SimPy lacked random-walk boundary clamping, SimPy's SIR helper hardcoded every agent as initially infected, Agents.jl used a hardcoded step count, SimPy, Melodie, and Agents.jl used a different movement kernel from AMBER, AgentPy, and Mesa, and Mesa's benchmark model did not route seeded runs through Mesa's model RNG. The corrected suite uses the same movement semantics and timing protocol across frameworks. The timings reported below reflect implementations that pass the relevant invariants before measurement.

\subsubsection{Variables and Measures}

The independent variables are framework (seven implementations as listed above), population size (500, 1,000, and 5,000 agents), and simulation length (50 steps). The dependent variable is execution time (wall-clock time for complete simulation, averaged over three runs with the slowest trimmed).

\subsubsection{Protocol}

Experiments were conducted on an Apple Silicon laptop running Python 3.12.7 and Julia 1.12.3. Framework versions were AMBER 0.3.1, Mesa 3.4.1, AgentPy 0.1.5, SimPy 4.1.1, Melodie 1.0, and Polars 1.32.2. All timings are wall-clock measurements. Each configuration was executed three times; the slowest sample was trimmed and the remaining two were averaged. The master runner at \path{benchmarks/run_all_frameworks.py} orchestrates every framework in one command.

\subsection{Results}

\subsubsection{Scaling Behavior}

\begin{figure}[H]
    \centering
    \includegraphics[width=\textwidth]{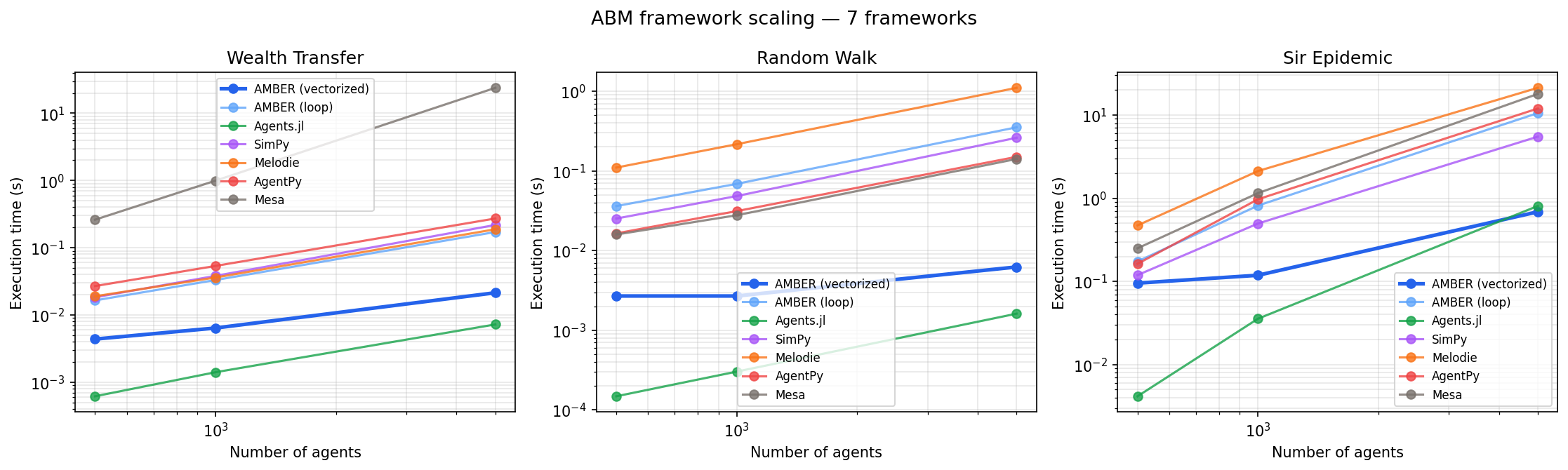}
    \caption{Execution time versus agent population size for three benchmark models and seven framework implementations. Lower values indicate faster execution. AMBER (vectorized) has the lowest execution time among Python-hosted implementations on every tested model; Agents.jl has the lowest execution time on the low-arithmetic-intensity wealth-transfer and random-walk kernels, while AMBER has the lowest execution time on the largest SIR workload.}
    \label{fig:scaling}
\end{figure}

Figure~\ref{fig:scaling} presents execution time as a function of population size across all three benchmarks and all seven frameworks, on log-log axes.

\subsubsection{Execution Time}

Table~\ref{tab:results} reports execution time at all three tested population sizes. Values are milliseconds unless otherwise noted; bold entries identify the lowest execution time for each benchmark and population size.

\begin{table}[H]
\centering
\caption{Execution time at 500, 1{,}000, and 5{,}000 agents over 50 simulation steps.}
\label{tab:results}
\small
\begin{tabular}{l rrr rrr rrr}
\toprule
 & \multicolumn{3}{c}{\textbf{Wealth Transfer}} & \multicolumn{3}{c}{\textbf{Random Walk}} & \multicolumn{3}{c}{\textbf{SIR Epidemic}} \\
\cmidrule(lr){2-4} \cmidrule(lr){5-7} \cmidrule(lr){8-10}
\textbf{Framework} & 500 & 1k & 5k & 500 & 1k & 5k & 500 & 1k & 5k \\
\midrule
AMBER (vectorized) & 4.4 & 6.4 & 21 & 2.7 & 2.7 & 6.2 & 95 & 119 & \textbf{687} \\
Agents.jl (Julia) & \textbf{0.6} & \textbf{1.4} & \textbf{7.2} & \textbf{0.1} & \textbf{0.3} & \textbf{1.6} & \textbf{4.2} & \textbf{36} & 808 \\
AMBER (loop) & 16 & 33 & 171 & 36 & 69 & 352 & 174 & 817 & 10{,}568 \\
SimPy & 18 & 38 & 218 & 25 & 48 & 261 & 119 & 497 & 5{,}506 \\
Melodie & 19 & 36 & 188 & 110 & 216 & 1{,}104 & 472 & 2{,}130 & 21{,}263 \\
AgentPy & 27 & 54 & 272 & 16 & 31 & 150 & 163 & 973 & 11{,}993 \\
Mesa & 261 & 996 & 23{,}917 & 16 & 28 & 141 & 250 & 1{,}150 & 17{,}967 \\
\bottomrule
\end{tabular}
\end{table}

\subsubsection{Speedup at 5{,}000 Agents}

Table~\ref{tab:speedups} re-expresses the 5{,}000-agent results as speedup ratios with AMBER (vectorized) as the baseline. Values greater than 1 indicate AMBER is faster; values less than 1 indicate the other framework is faster.

\begin{table}[H]
\centering
\caption{Speedup of AMBER (vectorized) vs each framework at $N = 5{,}000$ agents.}
\label{tab:speedups}
\begin{tabular}{lrrr}
\toprule
\textbf{Framework} & \textbf{Wealth Transfer} & \textbf{Random Walk} & \textbf{SIR Epidemic} \\
\midrule
Agents.jl (Julia)  & 0.3$\times$ (faster) & 0.3$\times$ (faster) & 1.2$\times$ \\
AMBER (loop)       & 8.0$\times$  & 56.8$\times$ & 15.4$\times$ \\
SimPy              & 10.2$\times$ & 42.0$\times$ & 8.0$\times$ \\
Melodie            & 8.8$\times$  & 178.0$\times$ & 30.9$\times$ \\
AgentPy            & 12.7$\times$ & 24.2$\times$ & 17.5$\times$ \\
Mesa               & 1117.6$\times$ & 22.8$\times$ & 26.1$\times$ \\
\bottomrule
\end{tabular}
\end{table}

\subsection{Analysis}

AMBER (vectorized) has the lowest execution time among Python-hosted frameworks on all three models and has the lowest execution time overall on the largest SIR benchmark. Agents.jl is faster on the two lower-cost microbenchmarks, where Julia's compiled dispatch has less fixed overhead than repeated Python-to-Polars calls. The results identify three performance regimes.

\textbf{Random Walk.} Agents.jl completes the 5{,}000-agent run in 1.6 ms compared with AMBER's 6.2 ms because the per-step body consists only of two independent coordinate updates and a boundary clamp. This leaves little computation over which Polars can amortize query planning and Python-to-Rust boundary costs. AMBER remains the lowest-latency Python-hosted implementation, roughly 23$\times$ faster than Mesa at 5{,}000 agents.

\textbf{SIR Epidemic.} AMBER completes the 5{,}000-agent run in 687 ms compared with 808 ms for Agents.jl (1.2$\times$ faster). The infection step is quadratic in population size because each infected agent must distance-test every susceptible agent. AMBER expresses this computation as a Polars cross join filtered by the squared-distance predicate, which runs as one compiled pipeline in Rust. Agents.jl performs the corresponding computation through specialized nested loops. At 5{,}000 agents the columnar pipeline offsets its setup cost; at smaller sizes Agents.jl benefits from lower per-call overhead.

\textbf{Wealth Transfer.} Agents.jl completes the 5{,}000-agent run in 7.2 ms compared with AMBER's 21 ms (3.0$\times$ faster). The per-step work is minimal: one array comparison, one random draw per active agent, and two scatter-add operations. In this regime, Polars' per-query overhead, including query planning, Python--Rust boundary crossings, and metadata bookkeeping, becomes a substantial component of total runtime. AMBER is nevertheless roughly $13\times$ faster than every other Python-hosted framework on this workload, and the AMBER loop path alone is 140$\times$ faster than Mesa.

\textbf{Object-oriented Python overhead.} Mesa's 5{,}000-agent wealth-transfer result (23.92 s) is 1118$\times$ slower than AMBER (vectorized) and 140$\times$ slower than AMBER (loop). This gap reflects Mesa 3.x's \texttt{AgentSet.shuffle\_do} materializing the full agent list per step for shuffle-ordered activation, compounded with scheduler-level per-agent dispatch overhead. The wealth-transfer workload stresses this pattern because every agent acts at every step.

\textbf{Limits of the columnar approach.} The wealth-transfer and random-walk results identify a regime in which a compiled JIT language has a structural advantage over a Python-hosted columnar framework. When the per-step work is $O(\mathrm{constant})$ per agent with little amortizable computation, SIMD vectorization alone does not eliminate Python-to-Rust call overhead. As the per-step computation becomes heavier, as in SIR, the columnar pipeline becomes competitive with compiled Julia while retaining a Python modelling interface.

\section{Reproducibility and Availability}

\subsection{Testing}

AMBER maintains an automated test suite of 245 unit and integration tests executed on every push to the \texttt{main} and \texttt{dev} branches via GitHub Actions. Coverage is tracked in CI, and tests are organised into:

\begin{itemize}
\item \textbf{Unit tests} (\path{tests/test_agent.py}, \path{tests/test_model.py}, \path{tests/test_sequences.py}, \path{tests/test_population.py}, and \path{tests/test_base.py}) covering the view API's attribute read/write protocol, the write-buffer flush semantics, \texttt{add\_agents} bulk creation, filtered and scatter view construction, and all legacy list-style compatibility surfaces.
\item \textbf{Environment tests} (\path{tests/test_environments.py}) covering grid, continuous-space, and network environments, including a regression guard for a torus-wrap neighbour-deduplication bug fixed as part of this release.
\item \textbf{Integration tests} (\path{tests/test_integration.py}) covering end-to-end simulation workflows. The file deliberately keeps two class suites: \texttt{TestFullSimulationWorkflows} exercises the legacy per-agent loop path as a backwards-compatibility regression guard, while \texttt{TestVectorizedWorkflows} exercises the view API idiom.
\item \textbf{Optimisation tests} covering the grid-, random-, Bayesian-, and SMAC-based calibration paths.
\end{itemize}

Continuous integration runs on Ubuntu, macOS, and Windows across Python 3.10, 3.11, and 3.12. The documentation is built with Sphinx under \texttt{-W --keep-going}, with warnings treated as errors, as part of the release workflow.

\subsection{Benchmark Correctness}

The benchmark suite includes a correctness stage that each framework implementation must pass before timing is considered. This stage executes each implementation against \path{benchmarks/correctness_check.py}, which checks per-model output invariants:

\begin{itemize}
\item \textbf{Wealth transfer}: total wealth must equal $N \cdot w_{init}$ at every step (Boltzmann conservation).
\item \textbf{Random walk}: all agent coordinates must lie within the configured world bounds.
\item \textbf{SIR epidemic}: the cardinalities $|S| + |I| + |R|$ must equal $N$ at every step, and the initial infected count must match the configured parameter.
\end{itemize}

This audit identified multiple benchmark defects: SimPy's wealth-transfer helper used a race-prone event update, SimPy lacked random-walk boundary clamping, SimPy's SIR helper hardcoded every agent as initially infected, Agents.jl used a hardcoded step count, SimPy, Melodie, and Agents.jl used a different movement kernel from AMBER, AgentPy, and Mesa, and Mesa's benchmark model did not route seeded runs through Mesa's model RNG. The corrected evaluation in Section~\ref{sec:evaluation} reflects implementations that pass the relevant invariants.

\subsection{Software Availability}

AMBER is released under the BSD 3-Clause licence and runs on Linux, macOS, and Windows with Python $\geq 3.10$. It requires Polars $\geq 1.0$, NumPy $\geq 1.20$, NetworkX $\geq 2.5$, Matplotlib $\geq 3.3$, Seaborn $\geq 0.11$, and scikit-optimize $\geq 0.9$. Optional dependencies include SMAC $\geq 2.0$ and ConfigSpace $\geq 0.7$ for advanced calibration, plus plotly and ipywidgets for interactive examples. No GPU is required.

The source repository is available at \url{https://github.com/a11to1n3/AMBER}; the version described here is \texttt{v0.3.1}. A Zenodo archive will be minted on acceptance through the GitHub--Zenodo integration. Anh-Duy Pham (University of W\"{u}rzburg) is the primary author and maintainer; contributor history is available at \url{https://github.com/a11to1n3/AMBER/graphs/contributors}. The software, user-facing documentation, comments, and tests are written in English.

\section{Discussion and Reuse Potential}

AMBER is intended as a general-purpose ABM framework for three overlapping groups of users.

\textbf{Domain scientists building mid-to-large-scale models.} AMBER is intended for epidemiological, ecological, economic, and social-simulation models whose population sizes reach the tens of thousands. Existing Mesa or AgentPy models can be ported with limited structural change: the \texttt{Model}/\texttt{Agent} class hierarchy is preserved, while population-wide updates move from per-agent loops to \texttt{agents.where(...)} expressions. The object-oriented path allows graph traversal, Schelling-style satisfaction checks, rule-based strategic behaviour, and other non-vectorised logic to coexist with the vectorised core.

\textbf{Method developers extending the columnar paradigm.} The \texttt{Population} class and the view-API base class (\texttt{\_BaseView}) are documented extension points. Researchers exploring GPU offload through cuDF, cluster-scale simulation through Arrow Flight, automatic-differentiation integration, or domain-specific view types can reuse AMBER's data model without reinventing state management. The three-view taxonomy (full, filtered, scatter) also leaves room for windowed views for time-series analyses and graph-neighbourhood views for networked models.

\textbf{Educators teaching ABM.} The legacy per-agent API remains supported and offers a familiar starting point for students new to agent-based modelling. Once a model is correct, the same class can be migrated incrementally to the view API as an example of vectorisation and performance engineering. The documentation at \url{https://amber.readthedocs.io} provides a quickstart tutorial, walkthroughs of three canonical models, and API reference material generated from source docstrings.

Support is provided through the GitHub issue tracker at \url{https://github.com/a11to1n3/AMBER/issues}. The repository includes a contributing guide describing the development workflow, review criteria, and expectations for bug reports, pull requests, and new examples.

\section{Limitations}

The evaluation is intentionally focused on three representative kernels rather than an exhaustive catalogue of ABM workloads. The benchmark suite covers population-wide state updates, continuous-space movement, and a spatial epidemic model, but it does not yet evaluate large graph dynamics, explicit GIS workloads, long-running calibration campaigns, or distributed simulation. The reported timings are also hardware-specific: all measurements were collected on one Apple Silicon laptop and should be interpreted as relative comparisons under the stated protocol rather than universal performance constants.

AMBER's columnar representation is most effective when model logic can be expressed as operations over many agents at once. Models dominated by branch-heavy per-agent reasoning, deep graph traversal, or small constant-time updates may benefit less from the columnar backend, and in some cases a compiled loop in a JIT language may remain faster. AMBER therefore complements rather than replaces compiled ABM frameworks: its primary contribution is to extend the range of model sizes that can be handled efficiently while remaining inside the Python scientific workflow.

\section{Future Directions}

The columnar architecture opens several directions for subsequent work. First, GPU-backed DataFrame systems such as cuDF could extend the same state representation to larger population sizes, provided that model updates can be expressed as GPU-friendly column operations. Second, static or dynamic analysis of model code could help identify agent loops that are mechanically transformable into view-based column updates. Third, integration with automatic-differentiation and probabilistic-programming tools could support gradient-informed calibration or likelihood-free inference workflows. Finally, distributed execution over partitioned columnar state could support simulations whose population or parameter-sweep scale exceeds a single workstation.

\section{Conclusion}

AMBER demonstrates that a Python ABM framework can retain familiar modelling abstractions while moving dominant population operations into a compiled columnar backend. The benchmark results identify both the strengths and limitations of this design: columnar execution provides large speedups over object-per-agent Python frameworks and can exceed the performance of a compiled Julia baseline on heavier spatial workloads, while very small per-agent kernels remain better suited to low-overhead compiled loops. This design provides a path toward scalable agent-based simulation within the Python scientific ecosystem.

\section*{Acknowledgements}

We gratefully acknowledge the open-source communities behind Polars and Apache Arrow, whose work on high-performance columnar data processing forms the foundation of this project.

\section*{Funding Statement}

This work received no specific grant from any funding agency in the public, commercial, or not-for-profit sectors.

\section*{Competing Interests}

The author declares no competing interests.

\bibliographystyle{plainnat}
\bibliography{paper}

\end{document}